\documentclass{article}
\usepackage{epsfig}
\usepackage{amsfonts}
\usepackage{amsmath}

\input{tcilatex}

\begin{document}
\title{\hspace{4.1in}\\
\vspace*{1cm}
{ Moduli and multi-field inflation}}
\author{\vspace*{0.5cm}
{ Zygmunt Lalak}\\
%\footnote{Based on a talk given at}\\
Institute of Theoretical Physics, University of Warsaw, Poland\\}
\date{ }
\maketitle

\begin{abstract}
Moduli with flat or run-away classical potentials  are generic in theories based on supersymmetry and extra dimensions. 
They mix between themselves and with matter fields in kinetic terms and in the nonperturbative superpotentials. 
As the result, interesting structure appears in the scalar potential which helps to stabilise and trap moduli and leads to 
multi-field inflation. The new and attractive feature of multi-inflationary setup are isocurvature perturbations which can 
modify in an interesting way the final spectrum of primordial fluctuations resulting from inflation. 
\end{abstract}
\vskip 1cm

\section{Introduction}

Last year's results reported by  WMAP3 \cite{Spergel:2006hy},\cite{Kinney:2006qm}
seem to confirm the inflationary paradigm with 
the spectral index $n_s < 1$ and agree with the cosmic concordance model strengthening the case for 
 dark energy. Therefore it is an actual  chalenge for the theory of fundamental interactions to accommodate 
any of the two ideas in realistic extensions of the Standard Model (BSMs). %In this note we discuss briefly a promising direction 
%of inflationary model building in the context of stringy supergravities, which belong to the best theoretically motivated BSM theories.

Four dimensional supergravities stemming from string theories posses field dependent couplings controlled by vevs of moduli - fields which 
have flat or featureless classical scalar potential. These fields interact with gravity-strength couplings and appear to be natural candidates to play the role 
of the inflaton. In addition, they mix at the level of kinetic or potential energy with all matter-like fields present in the Lagrangian, which makes them 
a part of the inflationary dynamics even if the designed would-be  inflaton is not a modulus but rather a matter field. Hence, the issue of generating supersymmetric 
inflation is intertwined with the issue of moduli stabilisation and it is a multi-field rather than a single field phenomenon, which may lead to important consequences. 
The presence of more than one active field makes the analysis of the dynamics more complicated. However, there are two crucial advantages of such an approach. 
Firstly, it becomes easier to model simultaneously succesful inflation, cancellation of the post-inflationary vacuum energy and trapping of the moduli. 
Secondly, the physics of inflationary fluctuations is markedly different in multi-field setups than in the standard single field inflationary models, which 
makes the realistic spectrum of fluctuations more natural and easier to generate.

\section{Modular inflation}

One interesting illustration of the possible role of additional scalars in generating realistic inflation is the 
case of the dilaton in heterotic string effective Lagrangian \cite{Lalak:2005hr}. It is well known that a pair of condensates (a racetrack) can easily generate 
a scalar potential with a nontrivial weakly coupled minimum for both scalar components of the dilaton superfield. 
An additional difficulty consists in the necessity to cancel the post-inflationary cosmological constant. To simplify the inflationary dynamics, let us assume that 
in addition to  the dilaton there exists another modulus, the volume modulus $T$, with a no-scale K\"ahler potential, which automatically makes the tree-level scalar
potential semi-positive definite. We assume in addition, that that no-scale modulus becomes stabilised in a different sector of the model, for instance with the 
help of non-trivial D-terms \cite{Lalak:2005hr}. Let us note at this point that the two pehnomena -- the cancellation of the cosmological constant and generation of the inflationary 
period are not independent, as discussed for instance in \cite{Ellis:2006ar}, but we ingnore this complication in what follows. 

The weakly-coupled racetrack cannot generate inflation. There are no regions where slow-roll parameters could be made sufficienly small simultaneously in
all directions in (two-dimensional) field space. Moreover, the exponential scalar potential is so steep along the real component  of the dilaton, that the field 
which starts its evolution from arbitrary initial conditions set at very high initial temperatures ``overshoots''  the shallow weakly-coupled minimum and 
runs away towards the ``cold'' asymptotic minimum at vanishing gauge coupling. 
Both problems can be solved if an additional superfield is present. In the case the mass, $m,$ has a dependence on the vacuum expectation value, 
$<\chi >,$ of a scalar field, $\chi ,$ it is clear that the condensation
scale and hence the dilaton potential which determines the condensation
scale will also depend on $<\chi >.$ This can arise through the moduli
dependence of the couplings involved in the mass generation after a stage of
symmetry breakdown. For example a gauge-nonsinglet field, $\Phi ,$ may get a
mass from a coupling to a field, $\Psi $, when it acquires a vacuum
expectation value through a Yukawa coupling in the superpotential of the
form $W=\lambda \Psi \Phi \Phi $. In
general the coupling $\lambda $ will depend on the complex structure moduli, 
$\lambda =\lambda (<\chi >)$ and so the mass of the $\Phi $ field will be
moduli dependent, $m=$\ $\lambda (<\chi >)<\Psi >.$ In what follows we will
consider the implications of a very simple dependence of $m$ on the 
\TEXTsymbol{<}$\chi >$ which is sufficient to illustrate how $\chi $ can
provide an inflaton if it is a (modulus) field which has no potential other
than that coming from the $m$ dependence of the condensation scale. 
%(and the above superpotential coupling). 
In particular we take $m=\alpha +\beta <\chi >$ where $\alpha $ is a mass
coming from another sector of the theory, possibly also through a stage of
symmetry breaking.  We shall assume for
simplicity a canonical kinetic term, $K=\bar{\chi}\chi ,$ for the modulus $%
\chi $, although in practice the form of its K\"{a}hler potential may be
more complicated.

We are now in a position to write down the form of the superpotential
corresponding to two gaugino condensates driven by two hidden sectors with
gauge group $SU(N_{1})$ and $SU(N_{1})$ respectively. For simplicity we
allow for a moduli dependence in the second condensate only. The race-track
superpotential has the form 
\begin{equation}
W_{npert}=\left (AN_{1}M^{3}e^{-S/N_{1}}-BN_{2}M^{3}e^{-S/N_{2}}\left( \frac{M^{2}}{%
(\alpha +\beta \chi )^{2}}\right) ^{\gamma} \right) \chi^p \, .
\label{spot1}
\end{equation}%
The coefficients $A,\,B$ are related to the remaining string thresholds $%
\Delta _{i}$ by $A=e^{-\Delta _{1}/(2N_{1})}/N_{1}$ etc. whose moduli
dependence we do not consider here, $\gamma$ is related to the difference of beta functions before and after a stage of symmetry breaking at the scale $m$.

We will now argue that the racetrack potential has all the
ingredients to meet the challenges posed by successful inflation. There are two main
aspects to this. Firstly the domain walls generated by the racetrack
potential naturally satisfy the conditions needed for topological
inflation.  As a result there will be eternal inflation within the
wall which sets the required initial conditions for a subsequent
period of slow-roll inflation during which the observed density
perturbations are generated. The second aspect is the existence of a
saddle point(s) close to which the potential is sufficiently flat to
allow for slow-roll inflation in the weak coupling domain. As noted
above this is not the case for the pure dilaton potential but does
occur when one includes a simple moduli dependence.

\subsection{Topological inflation}

As pointed out by Vilenkin \cite{Vilenkin:1994pv} and Linde \cite{Linde:1994wt}, ``topological inflation'' can occur within a
domain wall separating two distinct vacua. The condition for this to
happen is that the thickness of the wall should be larger than the
local horizon at the location of the top of the domain wall (we call
this the `coherence condition'). In this case the initial conditions
for slow-roll inflation are arranged by the dynamics of the domain
wall which align the field configuration within the wall to minimise
the overall energy. The formation of the domain wall is inevitable if
one assumes chaotic initial conditions which populate both distinct
vacuua and moreover walls extending over a horizon volume are
topologically stable. Although the core of the domain wall is stable
due to the wall dynamics and is eternally inflating, the region around
it is not. As a result there are continually produced regions of space
in which the field value is initially close to that at the centre of
the wall but which evolve to one or other of the two minima of the
potential. 
If the shape of the potential near the wall is almost flat
these regions will generate a further period of slow-roll inflation,
at which time density perturbations 
will be produced.

In the case of the racetrack potential the coherence condition
necessary for topological inflation appears to
be rather easily satisfied. This condition states that the physical
width of the approximate domain wall interpolating between the minimum
at infinity and the minimum corresponding to a finite coupling should
be larger than the local horizon computed at the location of the top
of the barrier that separates them. The width of the domain wall,
$\Delta$, is such that the gradient energy stored in the wall equals
its potential energy, $\left(\frac{2\delta}{\Delta}\right)^2 = V
(s_{\mathrm{max}})$, where $\delta = \log
(s_{\mathrm{max}}/s_{\mathrm{min}})$.~\footnote{One should use here
the canonically normalised variable $z = \log (s)$, $s$ being the real part of the dilaton.} 

The fact that the racetrack potential readily generates topological
inflation offers solutions to {\em all} the problems related to modular inflation. With chaotic initial conditions for the dilaton at the Planck
era the different vacuua will be populated because the height of the
domain walls separating the minima is greater than the equilibrium temperature $T_{\mathrm{eq}}$
so thermal effects will not have time to drive the dilaton to large
values. This avoids the initial thermal-roll problem. Then there will
be regions of space in which the dilaton rolls from the domain wall
value into the minimum at finite $s$, moving from larger to smaller
values, and thus avoiding the thermal-roll problem. In fact once
created the vacuum bag at finite $s$ is stable, because it cannot move
back into the core and over to the other vacuum --- the border of the
inflating wall escapes exponentially fast --- so the respective region
of space is trapped in the local vacuum. Finally, if after inflation
the reheat temperature is lower than the critical temperature
$T_{\mathrm{crit}}$~\footnote{Above $T_{\mathrm{crit}}$ the minimum may disappear due to thermal corrections in the hot gauge sector.} the thermal effects will be too small to fill in
the racetrack minimum at finite $s$ and the region of space in this
minimum will remain there, thus avoiding a late thermal-roll problem.

Although the existence of topological inflation seems necessary for a
viable inflationary model, by itself it is not sufficient to generate
acceptable density perturbations. What is needed is a subsequent
period of slow-roll inflation with the appropriate characteristics to
generate an universe of the right size ($>50-60$ e-folds of inflation)
and density perturbations of the magnitude observed. 
%In the next
%Section we show that the racetrack potential has the correct
%properties to achieve this if one includes a simple moduli dependence
%of the form discussed in Section~\ref{modulidependence}.

\subsection{Inflation in the weak coupling regime}

In this subsection we present an example of a viable inflationary model
in which the inflaton is the pseudo-Goldstone boson associated with
the phase $\theta $ of the field $\chi$.

\begin{figure}[tbh]
\begin{center}
\epsfig{file=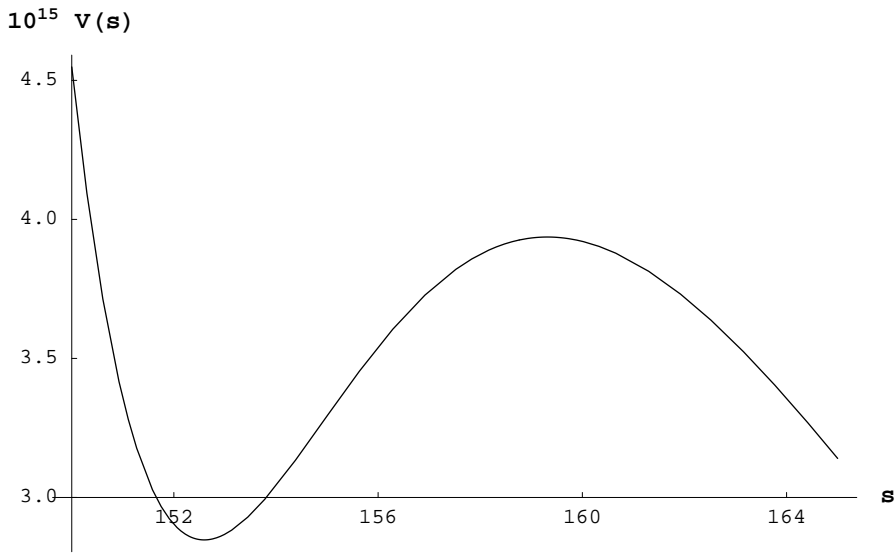, width=.7 \linewidth} 
\label{f2.1} %fig.3
\end{center}
\par
\caption{The $s$ dependence of the potential in the neighbourhood of
the weakly coupled minimum.}
\end{figure}

\begin{figure}[htb]
\begin{center}
\epsfig{file=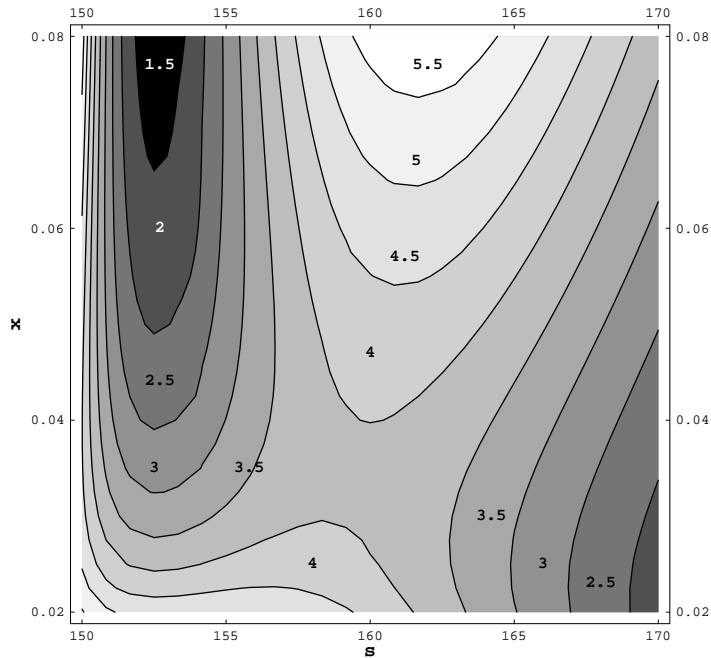, width=.55 \linewidth} 
\label{f2.2} %fig.4
\end{center}
\par
\caption{The contour plot in the $s,x$ plane in the neighbourhood of
the saddle point. The numbers on the contours multiplied by $10^{-15}$ give the corresponding vaules of the scalar potential in the units 
of $M^{4}_P$.}
\end{figure}

\begin{figure}[htb]
\begin{center}
\epsfig{file=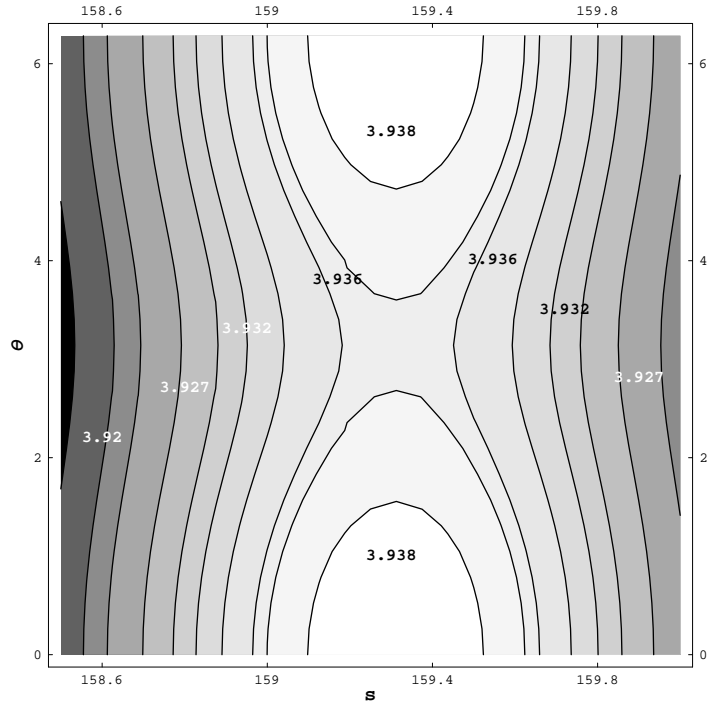, width=.55 \linewidth} 
\label{f2.3} %fig.5
\end{center}
\par
\caption{The $\protect\theta$ dependence of the potential in the
neighbourhood of the saddle point. Note that the slope {\em along} the
direction of $\theta$ is much smaller than the slope in the direction
of $s$. The numbers on the contours multiplied by $10^{-15}$ give the corresponding vaules of the scalar potential in the units 
of $M^{4}_P$.}
\end{figure}

%%%%%%%%%%%%%%%%%%%%%%%%%%%%%%%%%%%%%%%%%%%%%%%%%

Numerical analysis of the complete Lagrangian in the $(S,\,\chi )$
hyperplane shows that typically there are inflationary solutions. A
very nice example corresponds to the choice of parameters
$A=1.5,\;B=8.2,\;N_{1}=10,\;N_{2}=9,\;p=0.5,\;\alpha =1,\;\beta =2.3,$
and $\gamma =10^{-4} $. \ \ There is a weakly interacting minimum at
$s=152.6,\;\phi =0,\;x=0.42,$ and $\theta =3.16$ (we remind the reader
that $S=s+i\phi $ and $\chi =xe^{i\theta }$). \ The structure of the
potential in the neighbourhood of $s=152.6$ is shown in Figure 1 from
which it may be seen that there is a maximum at $s\approx160.$ There is a
domain wall between the weakly interacting minimum and the
non-interacting minimum at $s=\infty $. As may be seen from Figures 2
and 3 it has a saddle point at $s=162.2,$ $\phi =0,$ $x=0.074,$
$\theta =3.152.$ Inflation occurs within the domain wall and there is
further slow-roll inflation outside the wall as the wall inflates to a
size not supported by the dynamics generating the wall.

The initial conditions or this slow-roll deserves some comment. The
$s$ field starts very close to the saddle point at $s=162.2$. The same
is true for the fields $\phi $ and $x$ which, at the saddle point,
have masses larger than the Hubble expansion parameter at this
point. However the field $\theta $ is a pseudo-Goldstone field and
acquires a mass-squared proportional to $\gamma .$ As may be seen from
Figure 3, since $\gamma $ is small, the potential is very flat in the
$\theta $ direction and the mass of $\theta $ is much smaller than the
Hubble expansion parameter. As a result the vev of $\theta $ during
the eternal domain wall inflation undergoes a random walk about the
saddle point and so its initial value can be far from saddle point.

With these initial conditions it is now straightforward to determine
the nature of the inflationary period after the fields emerge from the
region of the domain wall. This corresponds to the roll of the fields,
$s$, $\phi$ and $x$ from the saddle point to the weakly interacting
minimum, but allows for $\theta $ to be far from the saddle point. The
$s,$ $\phi $ and $x$ fields rapidly roll to their minima. However the
gradient in the direction of the $\theta $ field is anomalously small
due to the pseudo-Goldstone nature of the field. Quantitatively, in
the neighbourhood of the weakly interacting minimum, we find a
negative eigenvalue of the squared mass matrix corresponding to the
phase $\theta $, and its absolute value is about $10^{4}$ times
smaller than the positive eigenvalues. This is much smaller than the
Hubble expansion parameter at the start of the roll and so the
$\theta$ field indeed generates slow-roll inflation. The remaining
degrees of freedom can be integrated out along the inflationary
trajectory. Inflation stops after about $7800$ e-folds at $\theta _{e}
= 3.54$ and the pivot point corresponds to $\theta _{\star} =
4.71$. The value of $\eta$ at this point is $\eta _{\star }=-0.0089$
so the the spectral index is $n_{\star }=0.98$, consistent with the
WMAP3 value at $2\sigma$. The agreement with the normalisation of the
spectrum is also readily achieved (we remind the reader that the
expectation value of $t$ can be considered as a free parameter for the
purpose of tuning the overall height of the inflationary potential, as
it is fixed in a separate sector of the model). Note that the
`running' of the spectral index is very small, $\mathrm{d}\ln
n_{\mathrm{s}}/\mathrm{d}\ln k<\times 10^{-5}$, hence probably
undetectable.

To summarise, the moduli dependent racetrack potential has a saddle
point which lead to a phenomenologically acceptable period of
slow-roll inflation with the inflaton being a component of the
moduli. No fine-tuning of parameters is required and the initial
conditions are set naturally by the first stage of topological
inflation. After inflation there will be a period of reheat and the
nature of this depends on the non-inflaton sector of the theory which
we have not specified here. From the point of view of the racetrack
potential the main constraint on this sector is that the reheat
temperature should be less than $T_{\mathrm{crit}}$ to avoid the
thermal roll problem. However $T_{\mathrm{crit}}$ is quite high, much
higher than the maximum reheat temperature allowed by considerations
of gravitino production, so we expect this constraint will be
comfortably satisfied in any acceptable reheating model.

\section{Inflation from matter  fields mixed with moduli}

The inflationary scheme described above encounters some problems when post-inflationary phenomenology 
is considered. First of all, the scale of the gravitino mass tends to be somewhat large with respect to the 
Fermi scale when the scale of inflation is made close to the upper limit imposed by COBE normalisation. 
Secondly, one might wish to rely on more traditional dynamics than topological inflation and topological trapping. 
In addition, in supersymmetric extensions of the Standard Model there are typically many quasi-flat directions involving 
combinations of scalars which belong to the matter rather than moduli sector. Hence, it is natural to consider matter-scalar driven 
inflation. However, as already explained in the introduction, and discussed at length in \cite{Ellis:2006ar}, matter scalars always mix with moduli. 
It turns out that very often the curvature along the direction of the relevant modulus is so large, that it spoils successful inflation 
along the direction of the matter scalar - the fields acquire large velocity transverse to the would-be inflationary trajectory and inflation ends prematurely. 
In the paper \cite{Ellis:2006ar} various ways out of this rather generic problem have been discussed. It turns out that sometimes certain amount of 
tuning and creative model building are necessary to generate sufficiently long epoch of slow-roll inflation. The discussion of this issue is rather model dependent,
and for more details and further references we recommend papers \cite{Ellis:2006ar} and \cite{Lalak:2007vi}.

\section{New features of multi-field inflation}

It is interesting to put aside fine details of the construction of multi-field inflationary models and ask for generic phenomena which occur in such a setup. 
The most interesting feature concerns quantum fluctuations during the inflationary period, which serve as primordial fluctuations triggereing 
formation of large scale structure. In the single-inflaton models fluctuations essentially do not evolve after horizon crossing, staying frozen until the second, post-inflationary 
horizon crossing, when the physical wavelength of a  fluctuation with a given wave number becomes smaller than the FRW  horizon.
However, when additional active inflatons are present, fluctuations in various fields get coupled and can evolve significantly after the first horizon crossing. 

\begin{figure}
\begin{center}
\hspace{-2.8cm}
\includegraphics*[height=6cm]{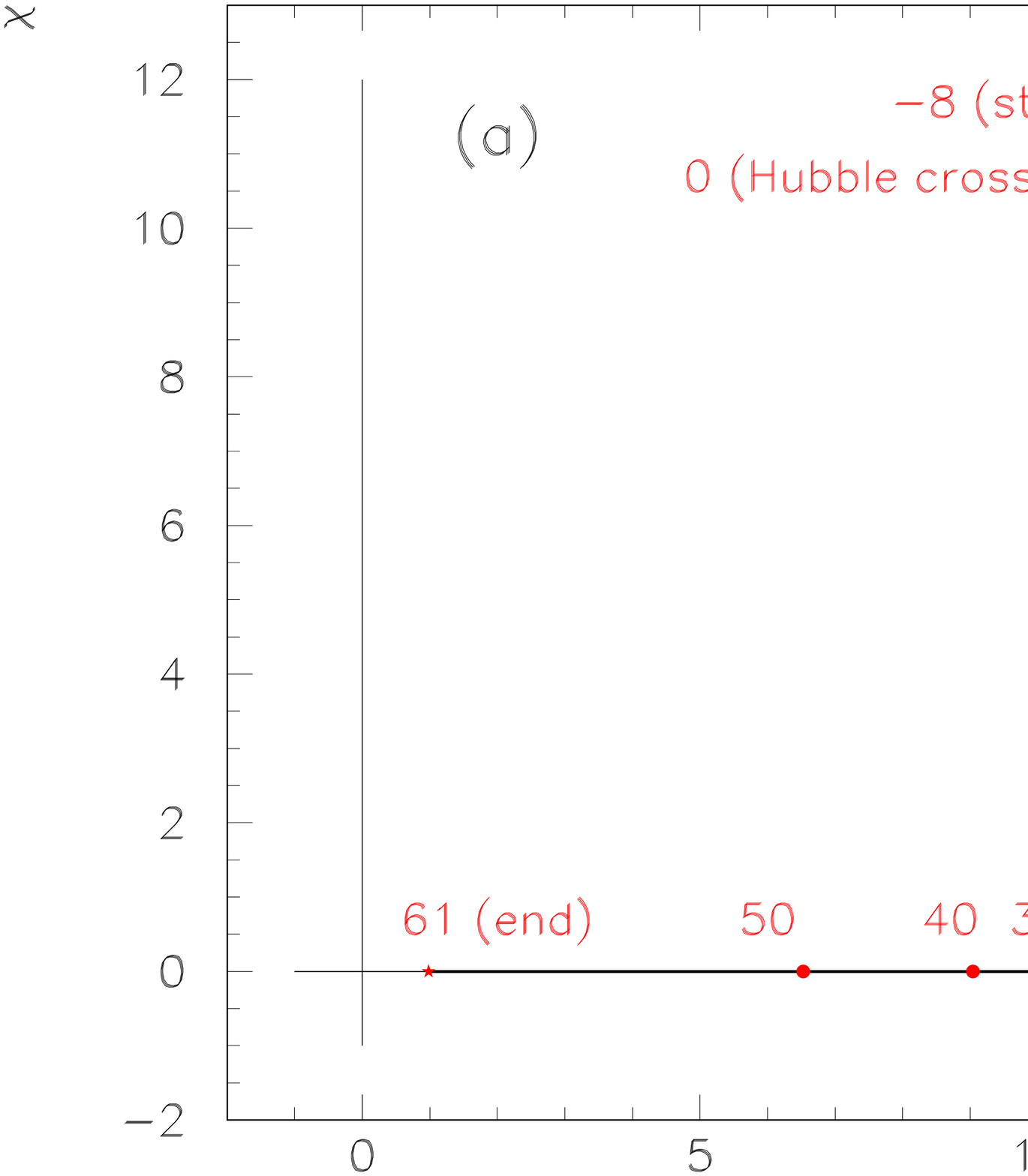}
\hspace{0.9cm}
\includegraphics*[height=6cm]{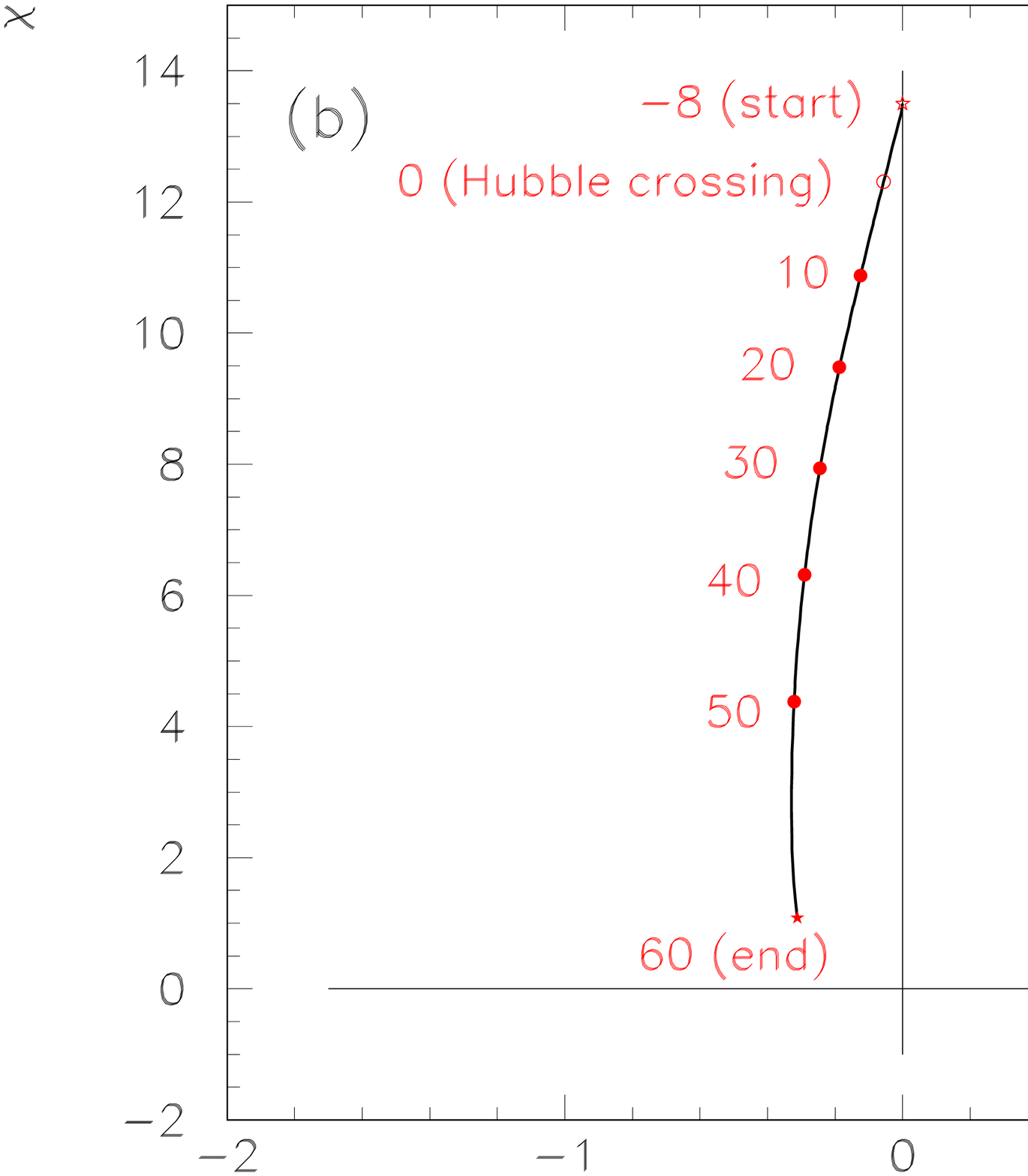}
\hspace{0.9cm}
\includegraphics*[height=6cm]{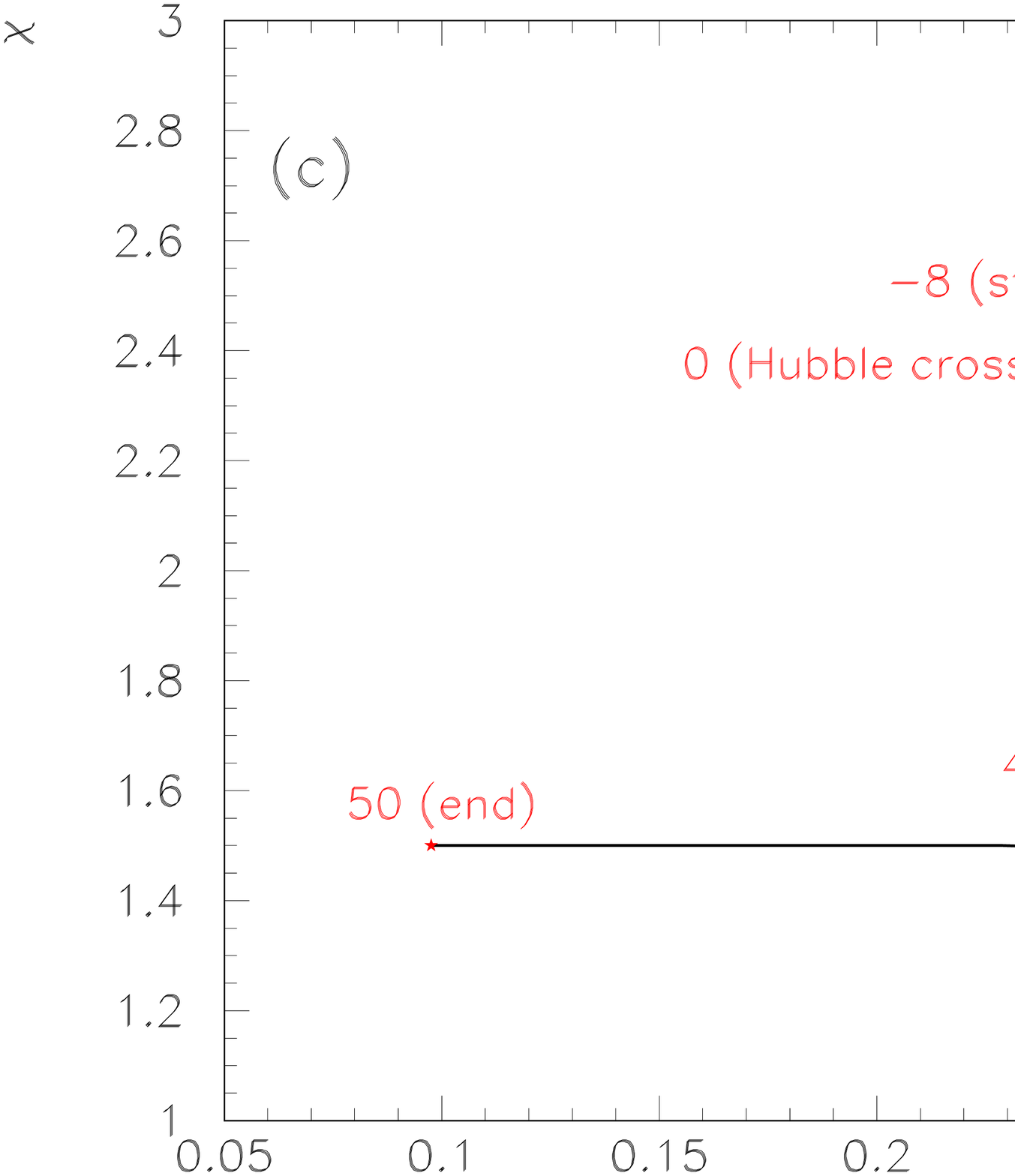}
\caption{\em Examples of classical inflationary trajectories for double 
inflation with canonical kinetic terms (left), double inflation with 
non-canonical kinetic terms (center) and roulette inflation (right).
The details of the models are described in  \cite{Lalak:2007vi}.
Subsequent
tens of efolds are indicated along the curves.\label{figcla}}
\end{center}
\end{figure}

In fact, this offers novel possibilities in shaping the spectrum of inflationary fluctuations. The point is that in the direction transverse to the classical inflationary trajectory 
the potential doesn't have to be as flat as along the classical trajectory. Hence, transverse fluctuations (isocurvature ones) do not need to have the spectrum as flat as 
that of the curvature fluctuations which borne as fluctuations in the `momentary' inflaton. If the coupling between both types of fluctuations is large enough, 
the isocurvature fluctuations can feed the curvature perturbations, and their not-so-flat spectrum becomes imprinted on the final spectrum of primordial fluctuations
resulting from inflation. As the result, it becomes reasonably easy to obtain the scalar spectral index significantly smaller than 1, as suggested by WMAP3.   
Examples of such solutions are presented in \cite{Lalak:2007vi}. Here we shall only illustrate the above statements. In the Figure \ref{figcla} examples of bent and 
curved classical inflationary 
trajectories are given. All of them give rise to significant isocurvature perturbations. 
To be more specific let us consider the so-called roulette inflation model which  has been investigated recently 
in \cite{Bond:2006nc} (see also \cite{Conlon:2005jm}). In the notation of \cite{Lalak:2007vi}
this model can be effectively described by
\begin{equation}
b(\phi) = b_0 - \frac{1}{3} \ln \left(\frac{\phi}{M_P}\right)
\end{equation}
and
\begin{equation}
V(\phi,\chi) = V_0 + V_1 \sqrt{\psi(\phi)}e^{-2\beta_1\psi(\phi)}+V_2\psi(\phi) e^{-\beta_1\psi(\phi)}\cos(\beta_2\chi) \, ,
\label{potbkpv}
\end{equation}
where 
$\psi(\phi)=(\phi/M_P)^{4/3}$ and
$b_0$, $V_i$, $\beta_i$ are functions of the parameters of
the underlying string model. A generic feature of the potential
(\ref{potbkpv}) is that it has an infinite number of minima arranged
periodically in $\chi$ and a plateau for large values of $\phi$,
admitting a large variety of inflationary trajectories, which may
end at different minima even if they originate from neighboring
points in the field space -- hence the model has been dubbed
{\em roulette inflation}. In this work, we adopt the parameter set
no.\ 1 (in Planck units: $b_0 = -11$; 
$V_0 = 9.0 \times 10^{-14}$;
$V_1 = 3.2 \times 10^{-4}$;
$V_2 = 1.1 \times 10^{-5}$; and
$\beta_1 = 9.4 \times 10^5$;
$\beta_2 = 2\pi/3$)
from \cite{Bond:2006nc} and choose the particular inflationary
trajectory shown in Figure \ref{figcla}. For this trajectory,
the factor $b_\phi M_P$ is rather large, of the order $10^3$, but
the effect of the non-canonical kinetic terms is strongly suppressed
by a very small value of $\epsilon$ on the plateau of the potential.
The smallness of $\epsilon$ also suppresses the energy scale of
inflation and one needs a smaller number of efolds than in the
models described above. For definiteness, we assumed that there
are $\sim50$ efolds between the moment that the scale of interest
crosses the Hubble radius and the end of inflation.

\begin{figure}[t]
\begin{center}
\hspace{-2.8cm}
%\includegraphics*[height=9cm]{fby6.ps}
%\hspace{1.5cm}
\includegraphics*[height=9cm]{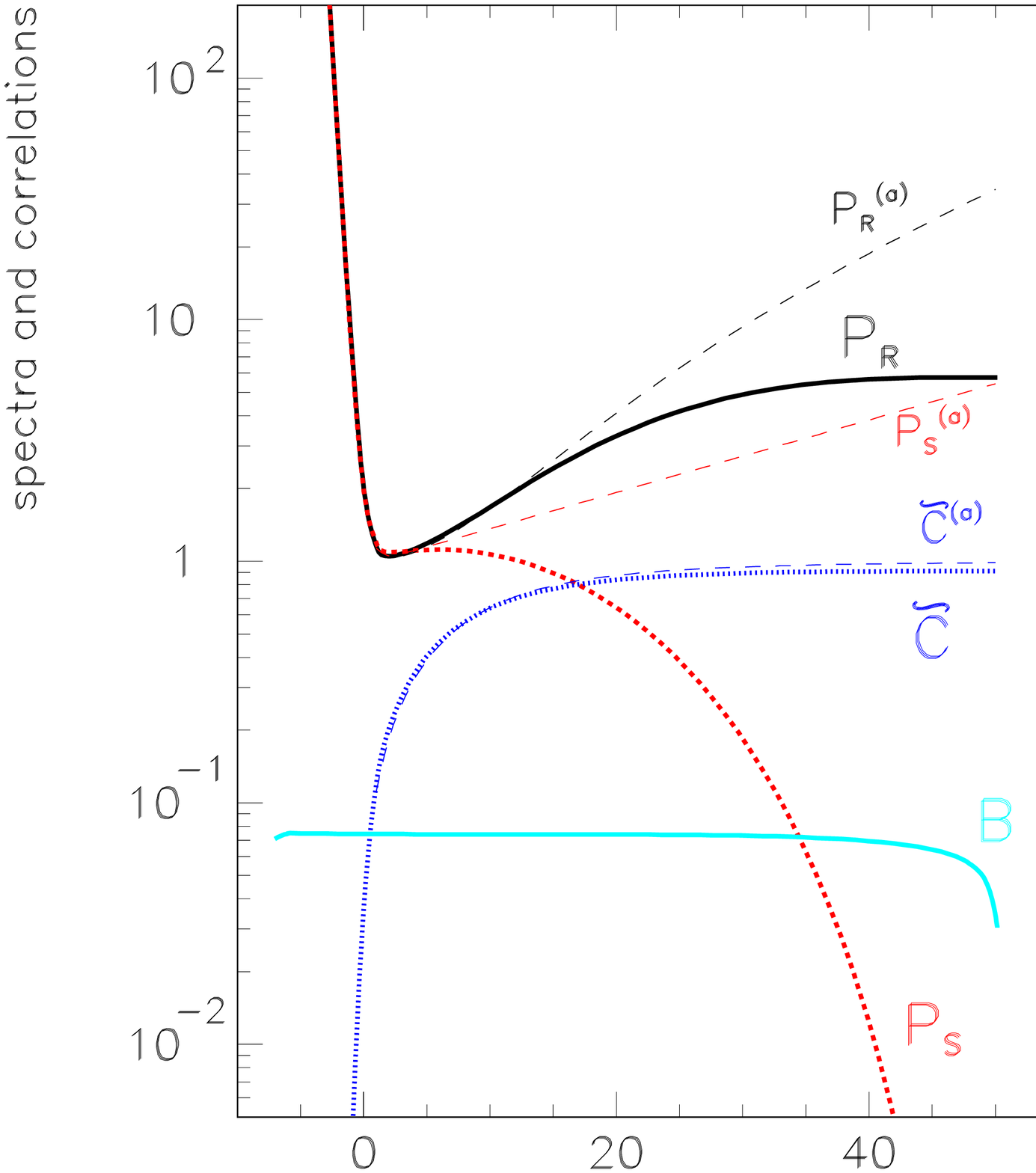}
\caption{\em 
Predictions for the spectra and correlations of the perturbations
in roulette inflation. 
Thick lines show the numerical results for ${\cal P}_{\cal R}$ -- power spectrum of curvature fluctuations, 
${\cal P}_{\cal S}$ -- power spectrum of isocurvature perturbations, 
normalized to the single-field result, respectively.
The coupling $B$ between the curvature
and isocurvature perturbations is also shown. (Full description can be found in \cite{Lalak:2007vi}.)
\label{figby}}
\end{center}
\end{figure}
The largest portion of the inflationary
trajectory in this example lies on the plateau of the potential
(\ref{potbkpv}), the slow-roll parameter $\epsilon$ is very small,
which makes the direct impact of the non-canonicality negligible.
The trajectory is, however, strongly curved in the field space
and the interaction between the isocurvature and curvature modes
is still important. As shown in \cite{Lalak:2007vi} one can accurately predict
the spectra and correlations in the vicinity of the Hubble crossing,
with deviations on super-Hubble scales resulting from the sourcing of
the curvature perturbations by the isocurvature ones.
Eventually, most of the curvature perturbations arise through this
effect.

\section{Summary and outlook}

%\vspace{1.5cm}
Multi-field inflation, with more than a single field active during the inflationary 
period, is a rather generic phenomenon in unification theories based on supersymmetry and 
extra dimensions. This fact has several implications for inflationary model building and also for the  
post-inflationary dynamics. 

First of all, in the case of modular inflation one can use some of the active  fields to trap run-away moduli 
via the topological inflation while the remaining ones could be used to create slow-roll inflation. 
A minimal example of such a  scheme presented here following \cite{Lalak:2005hr} has several attractive features:
\begin{itemize}
\item There is an initial period of topological inflation which sets the
initial conditions for slow roll inflation and avoids the rapid roll and
thermal roll problems usually associated with the racetrack potential. As a
result there is no difficulty in having our universe settle in the weakly
coupled minimum of the dilaton potential and not in the runaway
non-interacting minimum. 
\item The racetrack potential with simple moduli dependence has saddle
points which lead to slow roll inflation capable of generating the observed
density fluctuations with a spectral index smaller than but close to $1.$ Due 
to \ the
initial period of topological inflation the initial conditions for the slow
roll inflation are automatically set without fine tuning. 
\end{itemize}

Another far-reaching feature of multi-field inflation is the presence of isocurvature perturbations.  In some cases they may have a dramatic impact  on 
the final spectrum of primordial fluctuations which trigger structure formation. Quite often they easily lower the scalar spectral index  below 1, thus 
naturally reproducing the tendency implied by WMAP3 data. 

\vspace{0.4cm}

Based on talks given at {\em CTP Symposium on Supersymmetry at LHC} (Cairo, 
March 11-14 2007) and {\em String Phenomenology 2007} (Frascati, June 4-8 2007).
% AAAAAAAAAAAAAAAAAAAAAAAAAAAAAAAAAAAAAAAAAAAAAAAAAAAAAAAAAAAAAAAAAAAAAAAAAAAAAAAAAAAAAAAA

\vspace{0.5cm}

\centerline{\Large \bf Acknowledgements}

\vspace*{.3cm}

\noindent This work was partially supported by the EC 6th Framework
Programme MRTN-CT-2006-035863, and by the  grant MEiN  1P03D 014 26.
% \emph{The Quest for Unification: Theory
%Confronts Experiment}, 
%by Polish State Committee for Scientific Research
%grant KBN 1 P03D 014 26.

\vspace*{.5cm} 
%%%%%%%%%%%%%%%%%%%%%%%%%%%%%%%%%%%%%%%%%%%%%%%%%%%%%%%%%%%%%%%%%%%%%%%%%%%%%%%%%%%%%%%%%

\end{document}